# DEEP MULTI-TASK NETWORK FOR DELAY ESTIMATION AND ECHO CANCELLATION

*Yi Zhang, Chengyun Deng, Shiqian Ma, Yongtao Sha, Hui Song*

Didi Chuxing, Beijing, China

`{yizhang,dengchengyun,mashiqian,shayongtao,songhui}@didiglobal.com`

## Abstract

Echo path delay (or ref-delay) estimation is a big challenge in acoustic echo cancellation. Different devices may introduce various ref-delay in practice. Ref-delay inconsistency slows down the convergence of adaptive filters, and also degrades the performance of deep learning models due to 'unseen' ref-delays in the training set. In this paper, a multi-task network is proposed to address both ref-delay estimation and echo cancellation tasks. The proposed architecture consists of two convolutional recurrent networks (CRNNs) to estimate the echo and enhanced signals separately, as well as a fully-connected (FC) network to estimate the echo path delay. Echo signal is first predicted, and then is combined with reference signal together for delay estimation. At the end, delay compensated reference and microphone signals are used to predict the enhanced target signal. Experimental results suggest that the proposed method makes reliable delay estimation and outperforms the existing state-of-the-art solutions in inconsistent echo path delay scenarios, in terms of echo return loss enhancement (ERLE) and perceptual evaluation of speech quality (PESQ). Furthermore, a data augmentation method is studied to evaluate the model performance on different portion of synthetical data with artificially introduced ref-delay.

## 1. Introduction

Acoustic echo originates in a local audio loop back that occurs when a microphone picks up audio signals from a speaker, and sends it back to a far-end participant. Acoustic echo cancellation (AEC) or suppression (AES) aims to suppress echo from microphone signal whilst leaving the speech of near-end talker least distorted. Conventional echo cancellation algorithms estimate the echo path by using adaptive filters [1], under the assumption of a linear relationship between far-end signal and acoustic echo. In practice this linear assumption does not always hold, and thus a post-filter [2] [3] is often deployed to suppress the residue echo.

With the advancement in deep learning, many of the speech processing tasks, including acoustic echo cancellation, have been done using deep neural networks. Lee et al. [13] used a deep neural network with 3 layers of restricted Boltzmann machine (RBM) to predict the gain of residual echo suppression. Muller et al. [14] suggested to use a network of two FC layers to detect the activity of near-end signal. Zhang and Wang [15] proposed a bidirectional long-short term memory (BLSTM) to predict the ideal ratio mask from microphone signals. Carbajal et al. [16] built a two-layer network to predict phase sensitive filter of the residual echo suppression. Zhang et al. [17] used convolutional recurrent networks and long-short term memory to separate the near-end speech from the microphone recordings. Fazel et al. [18] proposed deep recurrent neural networks with multi-task learning to learn the auxiliary task of estimating the echo in order to improve the main task of estimating the near-end speech. Zhang et al. [19] proposed a generative adversarial network (GAN) with various metric loss functions to improve model robustness for both linear and nonlinear echoes.

Echo path delay estimation is crucial in echo cancellation, and AEC algorithms cannot work without a correct delay estimation. Lu et al. [4] proposed a light-computation-load algorithm by releasing input correlation and reducing cross-correlation lags. Govil [5] estimated the constant delay under the assumption that input sequence to the adaptive filter can be modeled as an autoregressive (AR) process whose order is much lower than the adaptive filter length. Sukkar [6] deployed a spectral similarity function based on cepstral correlation to detect acoustic echoes and estimate the echo path delay. To the best of the authors knowledge, deep learning based echo path delay estimation has not been well studied yet.

Direction of arrival (DOA) estimation task is kind similar to echo path delay estimation, which detects the source locations by estimating the time delay between microphones. This topic has been well studied in both conventional methods [7] [8] [9], as well as deep learning based methods [10] [11] [12]. Typically, the phase spectra are used as input for the deep learning models. However, DNN methods in DOA estimation cannot be applied to ref-delay estimation directly, since echo path delay could be much larger than that in DOA tasks, and thus delay information is not well kept in the phase spectra.

In this paper, we propose a multi-task network to estimate echo path delay and do echo cancellation simultaneously. The model consists of three subnets: two convolutional recurrent networks and one fully-connected network. CRNNs predict echo and target signals separately, and FC estimates the echo path delay. The multi-task model takes reference and microphone waveforms as input, and predicts the enhanced waveform as well as ref-delay as the output. We focus on the echo path delay effect on AEC performance in this paper, and thus other issues like nonlinearity are not covered here.

The remainder of this paper is organized as follows. Section 2 introduces the background knowledge. In Section 3 we present our multi-task algorithm, followed by experimental setting and results in Section 4. Final conclusion is given in Section 5.

## 2. Background knowledge

Acoustic echo is generated by the coupling of a microphone and a speaker, as shown in Figure 1. Far-end signal (or reference signal) $x(t)$ propagates from speaker and through various reflection paths $h(t)$, and mixes with near-end signal $s(t)$ to form the microphone signal $d(t)$. The acoustic echo is a modified version of $x(t)$ and includes echo path $h(t)$ and speaker distortion.

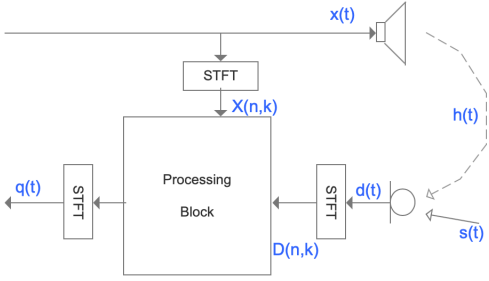

Figure 1 *Example of echo generation and acoustic echo cancellation*

Conventional AEC algorithms utilize adaptive filters to estimate the echo paths $h(t)$, and subtract the estimated echo $y(t) = \hat{h}(t) * x(t)$ from microphone signal $d(t)$. A separate double-talk detection is required to freeze filter adaption during double talk period. Often a post filter is needed to further suppress the residue echo.

Deep learning based AEC algorithms aim to find a mapping from input (echo corrupted signal) to output (target signal). With sufficient training data, neural network based solutions yield better performance than traditional ones in both matched and unmatched test cases.

For consistent ref-delay, both conventional and deep learning methods perform well. But in practice, echo path delay can be very different. For example, phones, iPads, and PCs all generate different ref-delays. Therefore, conventional methods often need a delay estimation component before the AEC block. For deep learning based methods, recordings collected from different devices may have very different ref-delays, and thus training set need to cover as many delay versions as possible to prevent mismatch in test set.

General metrics for AEC performance evaluation includes echo return loss enhancement (ERLE) and perceptual evaluation of speech quality (PESQ), which are used in this experiment as well.

ERLE is often used to measure the echo reduction achieve by the system during single-talk situation where the near-end talker is inactive. ERLE is defined as

$$ERLE(dB) = 10\log_{10} \frac{E\{d^2(t)\}}{E\{q^2(t)\}} \quad (1)$$

where $E\{\}$ represents the statistical expectation.

PESQ evaluates the perceptual quality of enhanced speech during double talk. PESQ score is calculated by comparing the enhanced signal to the ground-truth signal, its score ranges from -0.5 to 4.5 and a higher score indicates better quality.

## 3. Proposed method

The proposed algorithm consists of two convolutional recurrent networks and one fully-connected network, whose structure is shown in Figure 2.

The multi-task model takes microphone and reference waveforms as input, and predicts delay and enhanced waveforms as output. Input waveforms (dimension $T \times 1$, where T is waveform samples) is first fed into the encoder, which is the short-time Fourier transform (STFT) with window length 512 points and shift 256 points for 16000 Hz sampling rate, and output is the log-magnitude spectra ($K \times 257 \times 2$, where K is frame numbers). CRNN on the left in Figure 2 estimates echo signal from microphone waveform. Each CRNN includes three 2-D convolutional layers and three corresponding deconvolutional layers. Convolutions enforce the network to focus on temporally-close correlations in the input signal. Between the convolutional and deconvolutional layers there are two bidirectional LSTM layers to capture extra temporal information. Batch normalization (BN) [20] is applied after each (de-)convolutional layer except the last one. Exponential linear units (ELU) [21] are used as activation functions for each layer except the last layer that uses sigmoid activation function to predict the T-F masks. CRNNs also feature the skip connections [22], connecting each convolutional and corresponding deconvolutional layers. It passes the fine-grained information of the input spectra to the following layers.

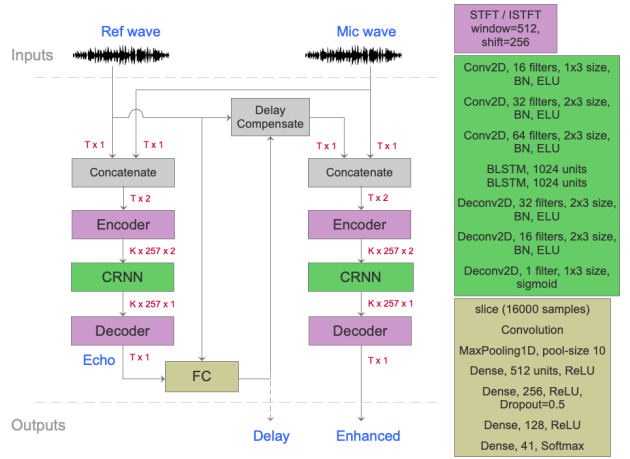

Figure 2 *Structure of the proposed multi-task network*

The estimated echo waveform, combined with reference signal, are fed into the fully-connected network. This network aims to predict the ref-delay based on the cross-correlation between echo and reference signals. The beginning 16000 samples from each waveform are used to compute the cross-correlation. This number is chosen based on our experiment, too small segment doesn't show clear correlation with the existence of various room impulse responses (RIRs), and too large segment provides no further improvement. Cross-correlation size is then reduced by maxpooling before entering the dense layer. It is shows that too large input size degrades the delay prediction accuracy in the experiment. Although the delay estimation resolution is declined by 10 dues to the maxpooling, CRNN is robust to handle this small variance, e.g., CRNN yields close results for ref-delay=10 and 19 samples (both will be categorized as 10 because of maxpooling). The three dense layers try to map the correlation input to a one-hot-vector output, where each spot represents a certain delay in samples. The RIRs used in this paper are collected from real environment, which have different shapes and peak locations (corresponding to the direct path). Hence, the existence of RIRs make it more challenge for the delay estimation, and thus fully-connected network shows more robustness than the '*argmax*' function. Furthermore, the ref-delay is generally much larger than that between microphones in DOA tasks, and thus phase spectra do not provide sufficient information for ref-delay estimation.

The estimated delay is used to compensate the reference signal. CRNN on the right in Figure 2 takes microphone and delay-compensated reference signal as input, and predicts the enhanced target signal. With the delay compensation, the 2nd CRNN model excludes the delay effect and focuses mainly on the RIR estimation, leading to better echo suppression and less target distortion.

The major model parameters are shown in Figure 2. Inside CRNN, the number of feature maps for convolutional layers are set to: 16, 32, and 64. The kernel size used for the first layer is (1, 3) and for the remaining layers is (2, 3), with strides set to (1, 2). The BLSTM layers consist of 2048 neurons, with 1024 in each direction and a time step of 100. Inside the FC network, convolution is conducted by '*tf.nn.convolution*'. Three hidden layers have 512, 256, and 128 units, with ReLU [23] as activation function. A dropout layer follows after the 2nd dense layer in order to prevent overfitting. The output layer has 41 units with Softmax activation function, indicating categories of ref-delay=0, 10, …, 400.

All models are trained using Adam optimizer [24] for 100 epochs with a learning rate of 0.0002 and a batch size of 1. The time step changes with the number of frames per sentence. The loss function for enhanced output is mean square error (MSE), and that of delay estimation is categorical cross-entropy, with a loss weight = [1, 1]. All the loss functions are based on utterance level.

Furthermore, the effect of data augmentation is studied. Synthetical data is generated by introducing delay into time aligned recordings. For example, we form a new training set that 20% of it is synthetical samples. The new training set allows model to learn delay prediction. Due to the imbalanced categories of ref-delays (e.g., 80% ref-delay=0), the categorical cross-entropy is replaced with focal loss [25] as the loss function. Loss weight remains unchanged from above.

## 4. Experimental evaluation

### 4.1. Experimental settings

TIMIT [26] is used to generate the dataset to evaluate the echo cancellation performance. We built a dataset similar to the ones reported in [19]: From 630 speakers of TIMIT, we randomly chose 100 pairs of speakers (40 male-female, 30 male-male, 30 female-female) as the far-end and near-end speakers. Three utterances of the same far-end speaker were randomly chosen and concatenated to create a far-end signal. Each utterance of a near-end speaker was then extended to the same size as that of the far-end signal by zero padding in the rear. Seven utterances of near-end speakers were used to generate 3500 training mixtures where each near-end signal was mixed with five different far-end signals. From the remaining 430 speakers, another 100 pairs of speakers were randomly picked as the far-end and near-end speakers. We followed the same procedure as described above, but this time only three utterances of near-end speakers were used to generate 300 testing mixtures where each near-end signal was mixed with one far-end signal. Therefore, the testing mixtures were from untrained speakers.

Five real environmental recorded RIRs from RWCP database [27] are used to generate acoustic echo in the experiment. Table 1 shows the information of the five RIRs.

Table 1 *RIRs from RWCP database*

| RIRs | E1A | E1B | E1C | E2A |
|---|---|---|---|---|
| $RT_{60}$ (in second) | 0.12 | 0.31 | 0.38 | 0.30 |

The main purpose of this paper is to study the delay estimation and its contribution to echo cancellation, and thus only linear echo scenarios are considered. In training step, microphone signals are generated randomly at signal to echo ratio (SER) {-6, -3, 0, 3, 6} dB, where SER is defined as

$$SER(dB) = 10\log_{10}\frac{E\{signal_{near}^2\}}{E\{signal_{far}^2\}} \quad (2)$$

In test stage, microphone signals are generated at SER levels {0, 3.5, 7} dB, slightly different from the training SERs, in order to evaluate the unmatched training-test cases.

Echo signal is simulated by convolving RIRs with far-end signals. Delay is randomly generated in a range [0, $D_{max}$], where $D_{max}$ is the upper limit for the ref-delay. We set $D_{max}$=400 samples in our experiment. Echo signal is then delayed and mixed with near-end signal under certain SERs.

### 4.2. Experimental results

In this experiment, two echo cancellation algorithms, EC-DE (echo cancellation & delay estimation) [4] and CRNN [17] are deployed as the benchmark. EC-DE is a conventional signal processing method adopting cross-correlation for delay estimation and linear adaptive filters for echo cancellation. CRNN is a deep learning based method and is identical to the CRNN subnet in the proposed method.

**A. Delay estimation**

We first evaluate the proposed method on delay estimation task in 'simple-delay' and 'RIR' scenarios. Near-end signal is inactive in this test, and microphone signal is obtained by convolving reference with a room impulse response. In 'simple-delay' case, microphone signal is simply a delayed version of reference signal, in other words, the room impulse response is a Dirac delta function. In 'RIR' case, RIRs from RWCP database are used to generate microphone signals. Delay is randomly generated between [0, $D_{max}$]. Figure 3 shows the five RIRs used in this experiment. These RIRs have different $RT_{60}$, as well as different peak locations. In the experiment, the RIRs are not time aligned in simulating echo signals.

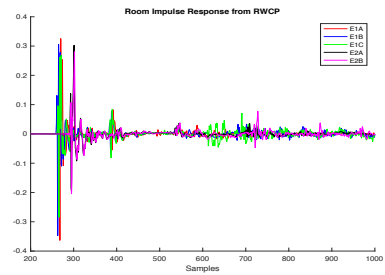

Figure 3 *Room impulse responses from RWCP dataset*

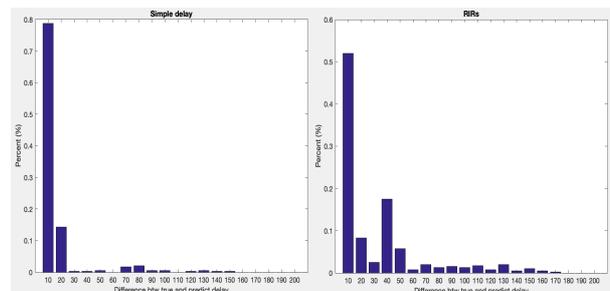

Figure 4 *Delay estimation for 'simple delay' and 'RIR'*

Figure 4 shows the delay estimation results for 'simple delay' and 'RIR' scenarios on 600 test sentences with SER 0dB. X-axis represents the difference between true and estimated delays in samples, and y-axis represents the percentage for each difference. For 'simple delay', 80% of the delay estimation difference is less than 10 samples. Accuracy of 'RIR' is a bit lower than 'simple delay', but still over 50% of the delay difference is less than 10 samples. The results show that the proposed method can effectively reduce the ref-delay variance by compensating the ref-delay, and potentially contribute to the echo cancellation performance.

**B. Echo cancellation**

In this test, we evaluate the echo cancellation performance of the proposed method and two benchmark methods. Two datasets 'set-A' and 'set-B' are generated. Echo signals are first obtained by convolving reference with RIRs. In 'set-A' echo is directly added to near-end signals to simulate microphone signals for each training (train-A) and test (test-A) mixture. In 'set-B' echo signal is randomly delayed between $[0, D_{max}]$ and then added to near-end signal for each training (train-B) and test (test-B) mixture. The PESQ and ERLE scores are shown in Table 2, which is averaged along different SERs. $D_{max}$ is not applicable to EC-DE since it has no training process.

Table 2 *PESQ and ERLE scores for train/test mismatch*

| Train/Test | | Train-A | | Train-B | |
|---|---|---|---|---|---|
| | | Test-A | Test-B | Test-A | Test-B |
| PESQ | EC-DE | 2.66 | 2.60 | 2.66 | 2.60 |
| | CRNN | 2.70 | 2.35 | 2.43 | 2.65 |
| | Multi-task | 2.72 | 2.38 | 2.70 | 2.75 |
| ERLE (dB) | EC-DE | 32.3 | 31.1 | 32.3 | 31.1 |
| | CRNN | 47.5 | 35.4 | 47.4 | 49.1 |
| | Multi-task | 48.0 | 35.9 | 50.4 | 50.9 |

CRNN yields good PESQ and ERLE scores in the matched cases (e.g., train-A & test-A, or train-B & test-B). However, its performance drops in the mismatched cases. For example, CRNN yields a PESQ score 2.35 for train-A & test-B, compared to 2.70 for train-A & test-A scenario, and the ERLE score decreases by 12.1dB (from 47.5 to 35.4). The proposed method shows no improvement in train-A & test-B case, since delay prediction model learns nothing from training and thus cannot contribute to the performance.

The proposed method yields better PESQ and ERLE scores than CRNN when training with set-B. It yields similar PESQ and ERLE scores for both matched and mismatched scenarios. But PESQ score of CRNN drops from 2.65 in matched case (train-B & test-B) to 2.43 in mismatched case, and ERLE score drops from 49.1 to 47.4. The results prove that delay prediction model compensates the delay and thus releases the mismatch effect.

EC-DE is robust to the delay since it has a delay estimation component inside, but its PESQ and ERLE scores are lower than the proposed methods as well.

**C. Data augmentation**

In the above section, we assume that either we collect all the data from certain devices (without extra delay, 'set-A'), or we collect the data from every possible device (without unexpected delay, 'set-B'). In practice our training set cannot cover all the delays, and thus data augmentation provides the diversity of delay for training models, without actually collecting new data.

In this experiment, we extend new training sets from 'set-A'. Synthetical data is generated by introducing random delay between $[0, D_{max}]$ into 'set-A' data. The new training set is formed where 20% or 50% of it is synthetical data. For 'set-B' in above section, each ref-delay category (0, 1, … 400) holds roughly 1/401 of total amount of data. However, if 20% of the training set is synthetical data, category of ref-delay=0 has 80% of the training data, and other ref-delay categories each has 0.05%. The imbalanced category issue severely degrades the model performance. Hence, focal loss function is used to replace cross-entropy in this test.

Table 3 gives the PESQ and ERLE scores where 20% and 50% portion of the training set is synthetical data. Note that, training portion=0% is identical to 'train-A', and portion=100% is identical to 'train-B' in section B.

Table 3 *PESQ and ERLE scores for augmented dataset*

| Training Portion | Test set | PESQ | | ERLE (dB) | |
|---|---|---|---|---|---|
| | | CRNN | MulTask | CRNN | MulTask |
| 0% | Test-A | 2.70 | 2.72 | 47.5 | 48.0 |
| | Test-B | 2.35 | 2.38 | 35.4 | 35.9 |
| 20% | Test-A | 2.68 | 2.71 | 46.2 | 49.3 |
| | Test-B | 2.42 | 2.68 | 37.1 | 48.9 |
| 50% | Test-A | 2.62 | 2.67 | 45.8 | 51.0 |
| | Test-B | 2.46 | 2.71 | 42.0 | 50.7 |
| 100% | Test-A | 2.43 | 2.70 | 47.4 | 50.4 |
| | Test-B | 2.65 | 2.75 | 49.1 | 50.9 |

With the portion of synthetical data increasing, CRNN yields smaller difference of PESQ and ERLE scores between match/mismatch cases. The proposed method obtains very similar scores for 'test-A' and 'test-B' over all the cases except portion=0%, indicating that 20% of augmented data is sufficient for delay estimation model. The proposed method achieves a 0.26 PESQ and 11.8dB ERLE improvement when 20% of training set is synthetical data ('test-B'), and 0.25 PESQ and 8.7dB ERLE improvement for 50% portion case ('test-B') over CRNN. Overall, the proposed multi-task model shows better robustness than CRNN to the inconsistent delay in the test.

## 5. Conclusion

In this study, we proposed a multi-task deep neural network to address the inconsistent delay in echo path. The proposed model consists of two convolutional recurrent networks and one fully-connected network, estimating echo signal, microphone signal, and echo path delay separately. Echo is estimated by the first CRNN and then is combined with reference signal to predict the echo path delay, finally the second CRNN takes delay compensated reference and microphone signals to estimate the enhanced target signal. By doing such, the proposed model gains robustness to the inconsistent ref-delay, and yields stable echo cancellation performance. Experiments indicate that the proposed multi-task model outperforms two benchmark methods under the criteria of PESQ and ERLE.